**Title:**
**Direct Bell states generation on a III-V semiconductor chip at room temperature**

**Authors:**
A. Orieux[1], A. Eckstein[1], A. Lemaître[2], P. Filloux[1], I. Favero[1], G. Leo[1], T. Coudreau[1], A. Keller[3], P. Milman[1], and S. Ducci[1]*

[1] Université Paris Diderot, Sorbonne Paris Cité, Laboratoire Matériaux et Phénomènes Quantiques, CNRS-UMR 7162, Case courrier 7021, 75205 Paris Cedex 13, France

[2] Laboratoire de Photonique et Nanostructures, CNRS-UPR20, Route de Nozay, 91460 Marcoussis, France

[3] Université Paris Sud, Institut des Sciences Moléculaires d'Orsay, CNRS - UMR 8214
Bâtiment 210 Université Paris-Sud 91405 Orsay Cedex

* corresponding author: sara.ducci@univ-paris-diderot.fr

**Abstract:**

We demonstrate the direct generation of polarization-entangled photon pairs at room temperature and telecom wavelength in a AlGaAs semiconductor waveguide. The source is based on spontaneous parametric down conversion with a counterpropagating phase-matching scheme. The quality of the two-photon state is assessed by the reconstruction of the density matrix giving a raw fidelity to a Bell state of 0.83; a theoretical model, taking into account the experimental parameters, provides ways to understand and control the amount of entanglement. Its compatibility with electrical injection, together with the high versatility of the generated two-photon state, make this source an attractive candidate for completely integrated quantum photonics devices.

PACS numbers: 03.67.Bg, 42.50.Dv, 42.82.-m, 42.65.Lm.

Entanglement [1] constitutes an essential resource for quantum information, speeding up algorithms [2,3], protecting encoded information from third party attacks [4] and enabling the teleportation of quantum states [5]. After a first generation of experiments aimed at demonstrating entanglement through the violation of Bell inequalities [6-8], a number of physical systems are under investigation for the development of future quantum technologies [9], and those involving photonic components are likely to play a central role [10,11,12,13]. Prior processes used to produce entangled photons were atomic radiative cascades [7] and, later on, spontaneous parametric down conversion (SPDC) in birefringent dielectric crystals [8]. The rapid development of quantum information and communication in the last two decades led to a demand for more practical sources of entangled photons, which could lead to commercial 'plug and play' devices. A significant effort has thus been devoted to explore the guided-wave regime, leading to entangled photon sources based on SPDC in periodically poled waveguides [14] and four-wave mixing (FWM) in optical fibers [15]; however, the dielectric nature of these platforms prevents a further progression towards monolithic sources. In this context, an attractive solution is provided by semiconductor materials, which exhibit a huge potential in terms of integration of novel optoelectronic devices; for instance, the bi-exciton cascade of a quantum dot has been used to demonstrate Bell state generation both in passive and active electrically-pumped configuration [16,17]. Unfortunately, these devices operate at cryogenic temperature, greatly limiting their potential for "plug and play" applications from a practical point of view. Moreover, they emit photons in a direction that is orthogonal to the wafer, whereas guided emission in its plane would simplify the implementation of a waveguide circuitry crucial for quantum information protocols on chip [10]. Very recently entangled photon generation has been reported on a silicon chip [18]: in that case entanglement is obtained by FWM, a third order process of lower efficiency that SPDC. Moreover the indirect band-gap of silicon hinders integration of the pump laser in the device.

More generally, the implementation of large-scale architectures for quantum information requires a minimal complexity for the building blocks. For this reason sources that do not require additional steps (such as walk-off compensation or interferometric schemes) to convert the correlated photons into highly entangled states, are desirable [19].

In this work we present a semiconductor source that addresses all these issues and directly generates polarization-entangled photons at room temperature in the telecom range by SPDC; this device can be considered the technological implementation of the EPR-Bohm gedanken experiment [20,21] since the entangled photons are emitted in opposite directions. The device is based on a specific phase-matching scheme [22] where a pump field ($\lambda_p = 759$ nm) impinging on top of a multilayer AlGaAs ridge waveguide with an angle $\theta$ generates two counterpropagating, orthogonally polarized waveguided twin photons ($\lambda_{s,i} = 1518$ nm) [23] (Fig. 1a,b).

As a consequence of the opposite propagation directions for the generated photons, two type II phase-matched processes occur simultaneously: the first one where the signal (s) photon is TE polarized and the idler (i) photon is TM, and the second one where the signal photon is TM and the idler one is TE. In the following, index (1) and (2) will refer to these two processes, and we will use the notation H for TE and V for TM. The central frequencies for signal ($\omega_s$) and idler ($\omega_i$) modes are determined through the conservation of energy and momentum in the waveguide direction, giving:

$$\omega_p = \omega_s + \omega_i,$$

$$\omega_p \sin\theta = \omega_s^{(1)} n_H(\omega_s^{(1)}) - \omega_i^{(1)} n_V(\omega_i^{(1)}),$$

$$\omega_p \sin\theta = \omega_s^{(2)} n_V(\omega_s^{(2)}) - \omega_i^{(2)} n_H(\omega_i^{(2)}),$$

where $n_H$ and $n_V$ are the effective indices of the guided modes, and $\omega_p$ is the pump frequency.

Momentum conservation in the epitaxial direction is satisfied by alternating AlGaAs layers with different aluminum concentrations (having nonlinear coefficients as different as possible) to implement a Quasi Phase Matching (QPM) scheme [23]. The inclusion of two Distributed Bragg Reflectors (DBRs) to create a microcavity for the pump beam results in an enhancement of the conversion efficiency and a reduction of the luminescence noise [24,25]. The sample was grown by molecular beam epitaxy on a (100) GaAs substrate. The epitaxial structure consists in 36-period asymmetrical $Al_{0.35}Ga_{0.65}As/Al_{0.90}Ga_{0.10}As$ distributed Bragg reflector (lower cladding + back mirror), 4.5-period $Al_{0.80}Ga_{0.20}As/Al_{0.25}Ga_{0.75}As$ quasi-phase-matching waveguide core and 14-period asymmetrical $Al_{0.35}Ga_{0.65}As/Al_{0.35}Ga_{0.65}As$ distributed Bragg reflector (as upper cladding + top mirror). The planar structure is chemically etched to obtain 2.5-3.5 μm thick ridges with 6-9 μm width.

The performances of this device can be expressed in terms of brightness: this figure of merit, corresponding to the rate of generated pairs per spectral bandwidth and pump power, is well suited to compare sources in different pumping regimes for integrated quantum photonics. Considering a typical average pump power of 1 mW, our device has a brightness about three orders of magnitude higher than that demonstrated in [18] thanks to the use of a second order nonlinear effect instead of a third order one.

In this work, the pump beam is provided by a pulsed Ti:Sa laser having a pulse duration of 3.5 ps, and a repetition rate reduced from 76 MHz to 100 kHz with a pulse picker. The mean power impinging on the waveguide surface is 1 mW. This pulsed regime can lead for instance to the control of frequency correlations in future experiments and take advantage of the best detection performances of the single photon avalanche photodiodes. In addition, we employed a technique to engineer the intensity profile of the pump beam to obtain polarization entanglement generation, which we will now describe.

The occurrence of the two simultaneous processes (1) and (2) allows to directly generate Bell states, as suggested by the tunability curves shown in Fig. 2. This graph shows that simultaneously pumping interaction 1 and 2 at the degeneracy angles $+\theta_{deg}$ and $-\theta_{deg}$ respectively generates two photons (from either interaction) with identical central frequencies. Filtering out the residual non-degenerate photons gives us, in principle, a maximally polarization-entangled photon state:

$$|\Psi^+\rangle = \frac{1}{\sqrt{2}}(|HV\rangle + |VH\rangle).$$

This experimental configuration can be realized by passing the pump beam through the center of a Fresnel biprism as shown in Fig. 1c. We used a glass biprism, each half of which deviates the pump beam entering the first face with a normal incidence by $\pm\theta_{deg}$. This peculiar property of our device, resulting from the transverse pump configuration and the type II phase matching, leads to entanglement generation within the device itself without the need of post manipulation of the generated pairs. This is a clear advantage allowing to reduce complexity and losses of large scale quantum photonic architectures.

The two-photon state was experimentally generated and then analyzed by a quantum tomographic reconstruction of the polarization state density matrix [26]. The emitted photon pairs are collected by two ×40 microscope objectives, analyzed in polarization with a quarter-wave plate, a half-wave plate and a polarizer, injected into single-mode fibers and detected by two InGaAs single-photon avalanche photodiodes with 25% detection efficiency and 500 ps gate, triggered by the pump laser. A time-to-digital converter (TDC) is used to measure the time-correlations between the counterpropagating photons. A long-pass interference filter and a fibred Fabry-Perot filter (1.2 nm FWHM) are used to filter out the residual luminescence noise and the non-degenerate pairs. Taking into account the overall transmission along the setup optical path, we estimate the total collection efficiency to 13%.

In Fig. 3, we show two of the sixteen coincidence histograms acquired for the tomography: one (HH) corresponds to coincidences between H-polarized photons (which can occur only because of noise), while the other (HV) corresponds to coincidences between orthogonally polarized photons. In our case, most of the coincidences are produced by the $\left|\Psi^+\right\rangle$ state.

The amount of noise in our measurements can be inferred from the polarization independent coincidences as presented in Fig. 3. The accidental coincidences amount to 0.04 Hz and are due in equal proportions to detector dark counts ($1.8\times10^{-4}$ per detection gate) and waveguide luminescence. We detect 0.77 Hz true coincidences, which, considering the detection (25%) and collection (13%) efficiency, corresponds to 0.007 pairs generated per pump pulse for each interaction with a repetition rate of 100 kHz. This signal to noise ratio limits the maximum measurable raw fidelity to 0.90 with our set-up. These figures could be improved by using detectors with higher detection efficiency and lower dark counts.

The raw density matrix (i.e. without background noise subtraction) reconstructed from our 16-measurement tomography [27] is presented in Fig. 4. Using these values, we can directly compute entanglement, estimated via the raw concurrence $C_{raw} = 0.68 \pm 0.07$ [28] and the raw fidelity $F_{raw} = 0.83 \pm 0.04$ to $\left|\Psi^+\right\rangle$. We have also tested the violation of a CHSH type inequality [29], which is often used as a benchmark to demonstrate the quality and usefulness of the entanglement generated from a quantum source. This test states that local realistic theories cannot provide a value larger than 2 for a combination of polarization correlation measurements: for the produced state we obtain the value of $2.23 \pm 0.11$. We have also reconstructed the so-called net density matrix, removing the accidental coincidences. We obtain a concurrence $C_{net} = 0.75 \pm 0.05$ and a fidelity $F_{net} = 0.87 \pm 0.03$ with the Bell state $\left|\Psi^+\right\rangle$.

The interpretation of these results can be done with the help of a simple model. Neglecting noise, the polarization density matrix of the generated two-photon state is [30]:

$$\rho = \alpha_1 |H,V\rangle\langle H,V| + \alpha_2 |V,H\rangle\langle V,H| + \beta |H,V\rangle\langle V,H| + \beta^* |V,H\rangle\langle H,V|,$$

where $\alpha_1$ (resp. $\alpha_2$) is the probability of pair generation through interaction 1 (resp. 2) by the pump beam impinging at $+\theta_{\text{deg}}$ (resp. $-\theta_{\text{deg}}$), with $\alpha_1 + \alpha_2 = 1$. $\beta$ quantifies the possible which-path information that disrupts the creation of maximally entangled photons. In our set-up, the two possible polarization states can be tagged by spectral and/or spatial degrees of freedom. The maximally entangled state $|\Psi^+\rangle$ is obtained for $\alpha_1 = \alpha_2 = 1/2$ and $\beta = 1/2$. In our case $\alpha_1 = \alpha_2 = 1/2$ and the spectral overlap is perfect when the angles of the two pumping beams are equal to $\pm\theta_{\text{deg}}$. This has been experimentally checked by measuring the spectra of the four emitted photons; all these spectra are superimposed within the resolution of our spectrometer (0.1 nm), indicating that the pumping angle is $\pm\theta_{\text{deg}}$ with a precision of 0.002°. We thus deduce that the parameter preventing $\beta$ to achieve its maximum value 1/2 is the spatial overlap of the two pumping beams which depends on their relative amplitude distribution profiles. As a first approximation, we have considered that, after passing through the biprism, each pumping beam is a half gaussian [30]. The point of maximum intensity of the two gaussians at the point they reach the sample is separated by a transverse distance of $\delta z$ leading to an imperfect overlap. Since $\beta$ is directly linked to the concurrence $C$ through $C = 2|\beta|$ [30], we can numerically investigate the impact of the experimental set-up alignment accuracy on the maximum entanglement level that can be measured (see Fig. 5). According to the theoretical model and the measured values of $C_{net}$ we deduce $\delta z = (0.3 \pm 0.1) \times w_p$ (where $w_p$ is the waist of the pumping beam impinging on the biprism equal to 2.4 mm in our case), a value in good agreement with the alignment precision of the present setup. Notice that, with the present shape of the pumping beams in the set-up, $C_{net}$ saturates for $\delta z = 0$ at a value of 0.84.

The results of Fig. 5, together with the theoretical analysis thus illustrate the important role played by the spatial overlap of the pumping beams and provide us with a clear guideline for controlling the degree and the purity of entanglement in polarization. It is worth noticing that the previous analysis could be applied to an, up to now, unexploited aspect of the present source: the interplay between the different photonic degrees of freedom (spatial, frequency and polarization) can be used to engineer hybrid hyper-entangled states of polarization and frequency degrees of freedom. This continuous-discrete entangled state has several applications in quantum information theory, as for instance, in quantum key distribution [31].

In conclusion, we have demonstrated what is to our knowledge the first III-V semiconductor source of polarization-entangled photon pairs working at room temperature and at telecom wavelength. The counterpropagating geometry allows a direct generation of entangled states that violate Bell type inequalities without the need of additional steps to remove which-way information, and the model proposed in this work describes well the role of the pumping configuration and device parameters, connecting them to the quality of the produced entangled state. The value of the fidelity demonstrated in this paper is comparable with the state of the art of the semiconductor sources while taking benefit of the strong potential of integration of the AlGaAs platform. For example, our source can be further evolved and miniaturized through the integration of an electrically-pumped vertical cavity surface-emitting laser acting as pump beam whose light could be directly coupled with any required angle and shape through an integrated diffraction grating. As shown by our model, a better spatial shaping of the pump would allow reaching even higher values of fidelity.

These results pave the way towards the demonstration of other unique properties associated to the counterpropagating geometry, such as the control of the frequency correlation nature via the spatial and spectral properties of the pump beam [30, 32] leading to extremely versatile hyperentangled states. Finally, its compatibility with the telecom network makes the device presented in this work an attractive candidate for scalable photonics-based quantum computation and quantum communications protocols.

**Acknowledgments:**

This work was partly supported by the French Brazilian ANR HIDE project and by Région Ile de-France in the framework of C'Nano IdF with the TWILIGHT project and SESAME Project 'Communications quantiques'. We acknowledge G. Boucher for help with experiments and S. Tanzilli for fruitful discussions. S.D. is member of Institut Universitaire de France.

**Figure Captions:**

FIG 1: (color online) Operation principle of the device

a,b : Counterpropagating phase-matching scheme. a: Interaction 1 with a pump angle $+\theta_{deg}$ produces a H-polarized signal and V-polarized idler; b: Interaction 2 with a pump angle $-\theta_{deg}$ produces a V-polarized signal and H-polarized idler. In both cases we call signal (idler) the photon exiting from the right (left) facet. Phase-matching is obtained automatically along the z direction and through a periodic modulation of the waveguide core in the x direction. The cladding consists of two Bragg reflectors to enhance the conversion efficiency. c: Sketch of the experimental set-up to generate Bell states: a laser pump beam impinges on a ridge waveguide microcavity with two symmetrical angles of incidence $+\theta_{deg}$ and $-\theta_{deg}$, through a Fresnel biprism, and entangled pairs of counterpropagating photons are emitted and collected at both facets. The waveguide is illuminated on its whole length ($L = 1.8$ mm).

FIG 2: (color online) Tunability curves

Simulated signal and idler tuning curves as a function of the angle of incidence of the pump beam, for $\lambda_p = 759$ nm. Polarization entangled photons can be generated by simultaneously pumping interaction 1 at $+\theta_{deg}$ and interaction 2 at $-\theta_{deg}$.

FIG 3: (color online) Photon correlation measurements

Time-coincidence detection histogram of signal and idler photons having respective polarizations H,V and H,H. Most of the coincidences are produced by the entangled state, whereas a small part of them stems from noise. Since noise is polarization independent, its contribution to HV coincidences can be inferred from the amount of HH coincidences.

FIG 4: (color online) Polarization state tomography

Real and imaginary part of the density matrix $\rho$ of the two-photon state, as reconstructed from quantum-state tomography.

FIG 5: (color online) Dependence of the net concurrence on the pump angle and the spatial overlap between the two pumping beams

The contour plot results from numerical simulation, while the gray ellipses correspond to the experimental results. The fact that the net concurrence saturates to a value of 0.84 when $\delta z = 0$ and the pump angle is equal to $\theta_{deg} = 0.35°$ is due to the particular shape of the pumping beams [30].

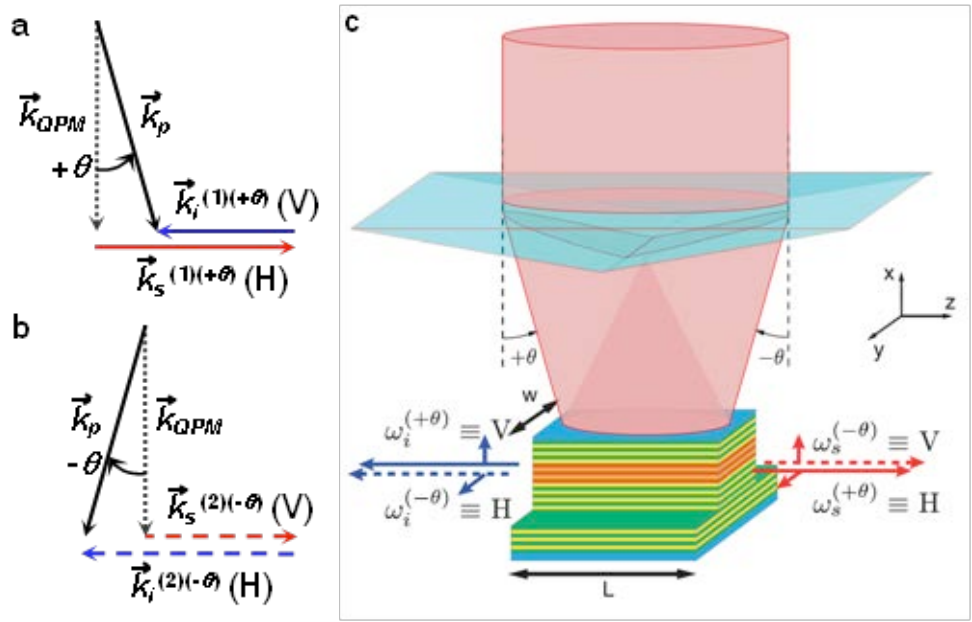

Figure 1

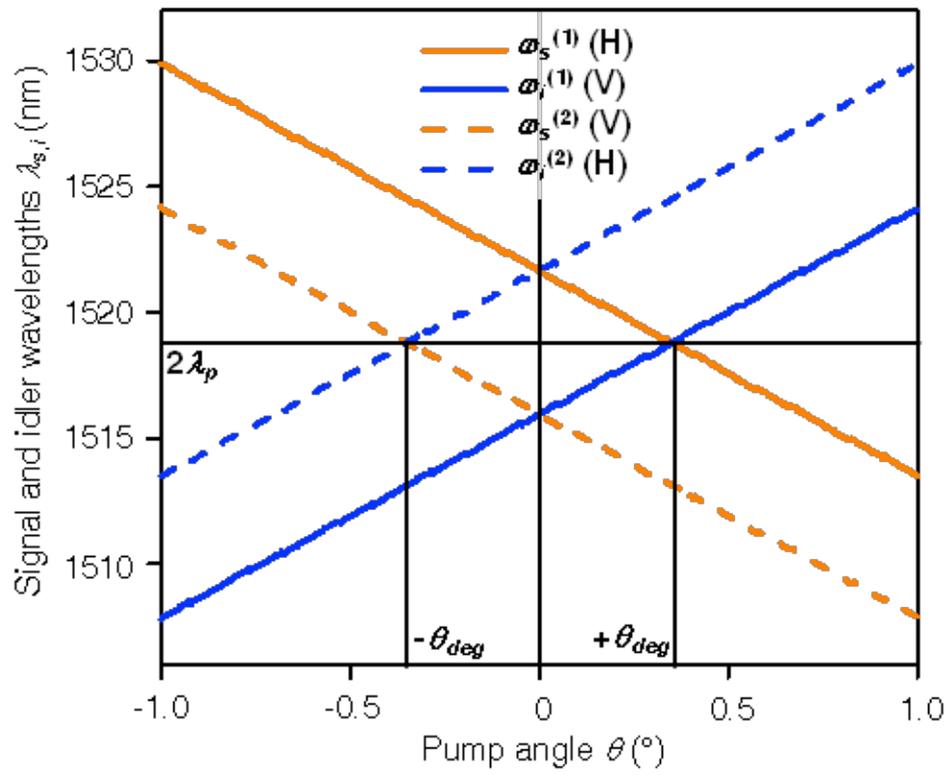

Figure 2

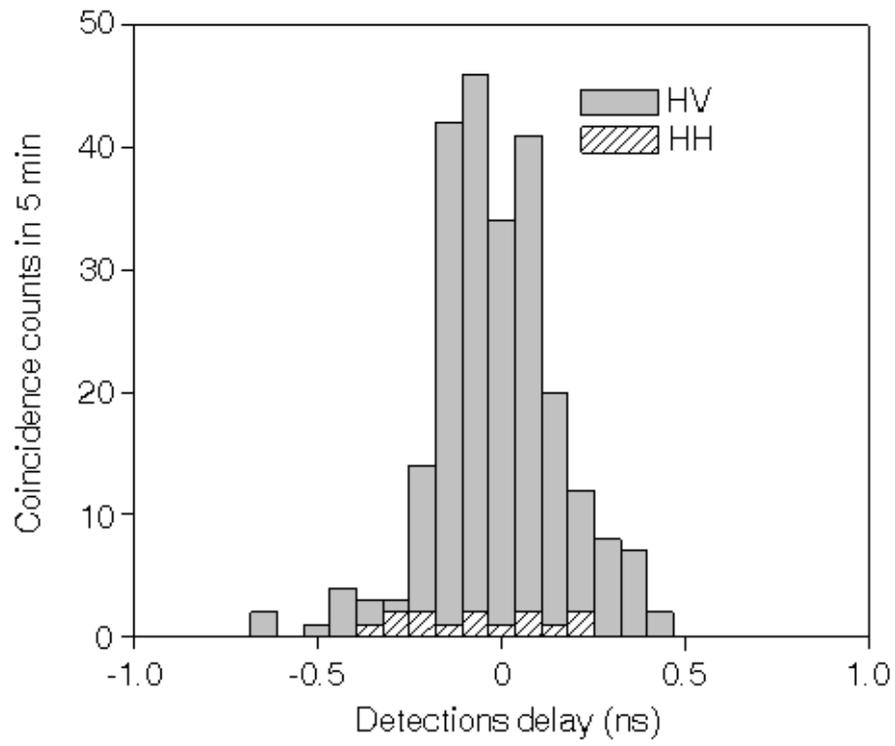

Figure 3

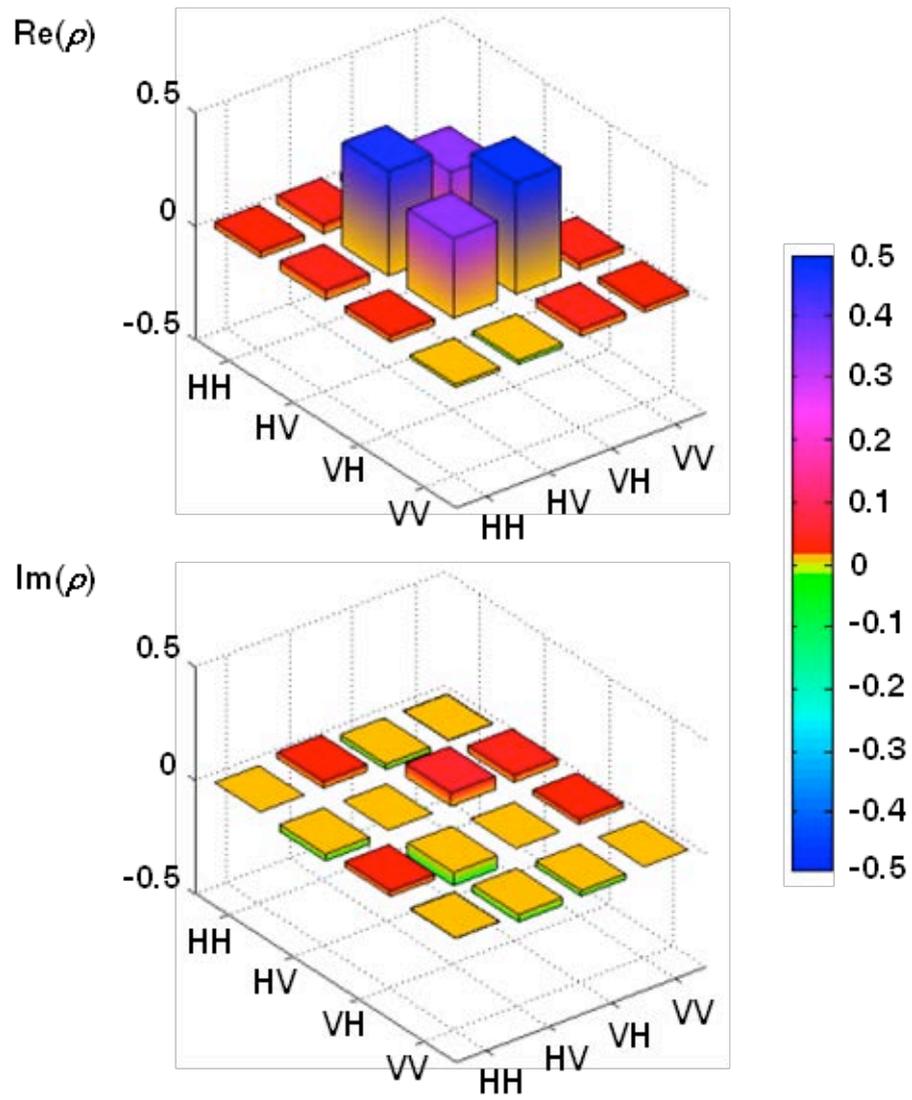

Figure 4

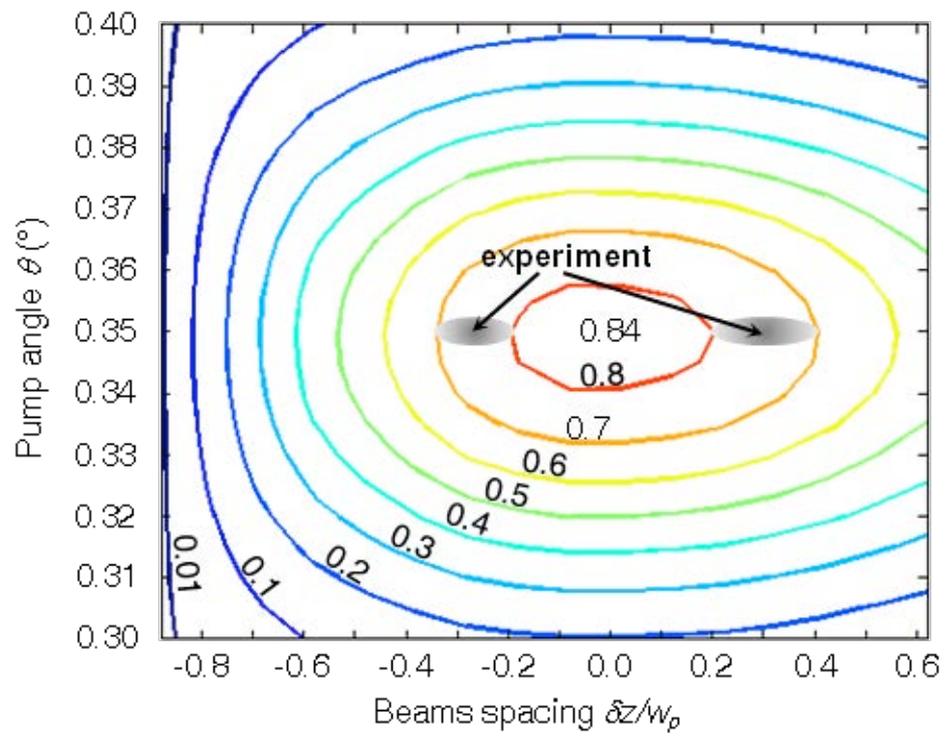

Figure 5